\pdfoutput=1
\documentclass[10pt,letterpaper]{article}
\usepackage{opex3}
\usepackage{cite}
\begin{document}

\title{­­­­­­­­Terahertz emission by diffusion of carriers and metal-mask dipole inhibition of radiation}

\author{M. E. Barnes,$^1$ D. McBryde,$^1$ G. J. Daniell,$^1$ G. Whitworth,$^1$ A. L. Chung,$^1$ A. H. Quarterman,$^1$ K. G. Wilcox,$^1$ A. Brewer,$^2$ H. E. Beere,$^2$ D. A. Ritchie,$^2$ and V. Apostolopoulos$^{1*}$} 

\address{$^1$School of Physics and Astronomy, University of Southampton, Southampton, SO17 1BJ, United Kingdom\\
$^2$Semiconductor Physics Group, Cavendish Laboratory, University of Cambridge, Cambridge CB3 0HE, United Kingdom}

\email{$^*$v.apostolopoulos@soton.ac.uk}

\begin{abstract}
Terahertz (THz) radiation can be generated by ultrafast photo-excitation of carriers in a semiconductor partly masked by a gold surface. A simulation of the effect taking into account the diffusion of carriers and the electric field shows that the total net current is approximately zero and cannot account for the THz radiation. Finite ele­­ment modelling and analytic calculations indicate that the THz emission arises because the metal inhibits the radiation from part of the dipole population, thus creating an asymmetry and therefore a net current. Experimental investigations confirm the simulations and show that metal-mask dipole inhibition can be used to create THz emitters.
\end{abstract}

\ocis{(320.7130) Ultrafast processes in condensed matter, including semiconductors;
(300.6495) Spectroscopy, terahertz.}

\section{Introduction}
\begin{figure}[tb]
\centering
 \includegraphics[scale=0.65]{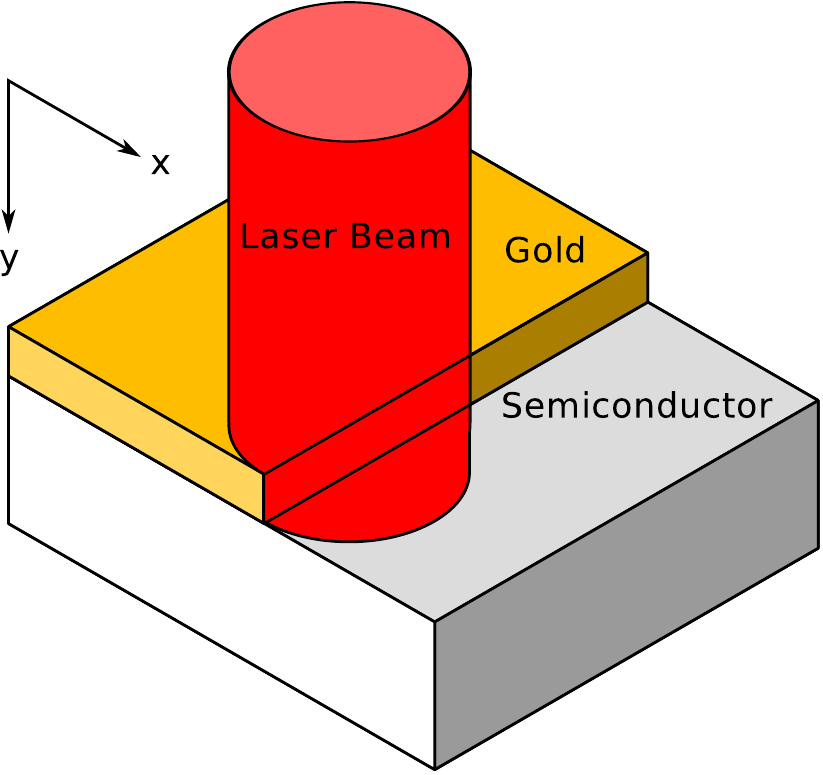}
 \caption{\label{fig:1} Illustration of the LPD effect showing how the laser beam is partly masked by a metallic region.}
 \end{figure}

\begin{figure}[htb]
\centering
 \includegraphics[scale=0.25]{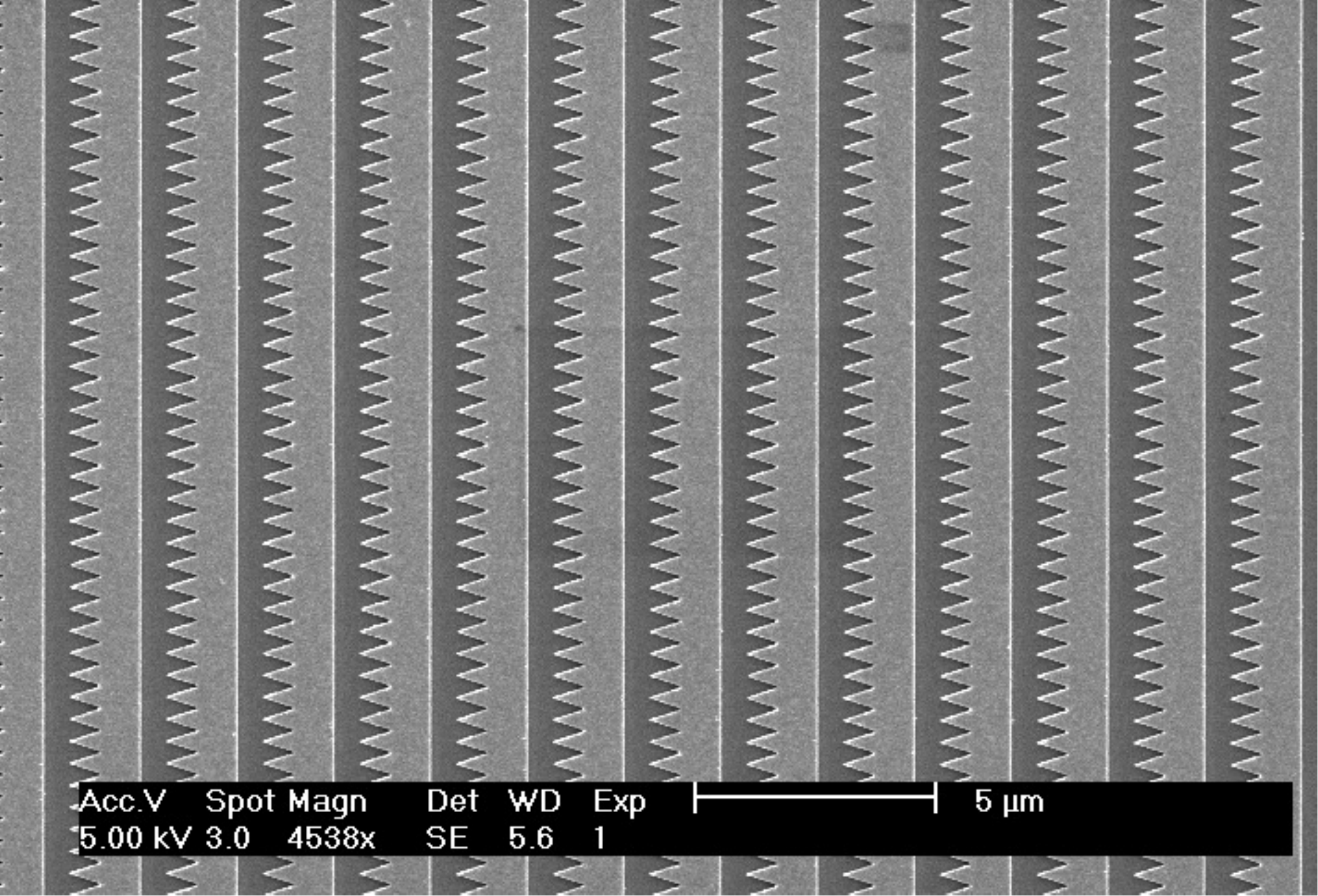}
 \caption{\label{fig:2} SEM of  a 2D multiplex emitter based on the concepts given in \cite{Klatt2010}  which failed to produce any measurable THz emission.}
 \end{figure}

THz time domain spectroscopy (THz-TDS) employs a synchronous detection scheme with inherent high signal to noise ratio which allows very accurate measurements. In THz-TDS a gallium arsenide photo-conductive (PC) emitter antenna is the most frequently used way of producing radiation \cite{Jepsen2011}. A titanium sapphire (Ti:S) ultrafast laser is commonly used to excite carriers across the band-gap of gallium arsenide. The pulsed illumination from the laser combined with an applied electric bias creates a fast current change, which emits THz waves with intensity proportional to the rate of change of current. The photo-Dember (PD) effect is another THz emission mechanism also based on ultrafast carrier transport. THz radiation is produced by illumination of a semiconductor surface by an ultrafast near infrared (NIR) laser with energy above the bandgap (usually Ti:S), but without the application of electrical bias \cite{Johnston2002,Liu2006,Gu2002}. The strong absorption of light near the surface creates a large carrier gradient of electrons and holes, which initiates a diffusion current. Because of the different mobilities, electrons and holes spatially separate on a picosecond time scale. The resulting dipole radiates at THz frequencies but in a direction perpendicular to the optical illumination. The PD effect was not considered to be competitive with other generation techniques because of low output power, mainly due to poor out-coupling \cite{Johnston2002,Liu2006,Gu2002,Johnston2002a}. However, Klatt et al. \cite{Klatt2011,Klatt2010,Klatt2010a} have demonstrated a lateral PD (LPD) emitter, shown in Fig.\ \ref{fig:1}(a), which takes advantage of the dipole created by lateral diffusion currents created by metal masking. This geometry exhibited bandwidth comparable to PC antennae and a series of emitters was fabricated demonstrating comparable power output to a PC antenna \cite{Klatt2010}. \\
We envisaged an array of emitters with the design shown in Fig.\ \ref{fig:2} and based on the principle given in \cite{Klatt2010} in order to improve ease of fabrication and performance. The LPD effect for these emitters was simulated using a 2 dimensional simulation of the diffusion and drift current and we approximated the THz emission using the derivative of the current density in \cite{McBryde2011}. However, further simulations for the proposed geometry indicated that the predicted THz radiation was much weaker than expected\cite{arxiv}. Here we report that the emitters were fabricated, as shown in Fig.\ \ref{fig:2}, and measurements confirmed that they generate no measurable THz signal. This result conflicts with the current understanding of the LPD effect, as outlined in \cite{Klatt2010} and the fact that the emitters of \cite{Klatt2010} do produce measurable signals. \\
 In this paper, to understand this discrepancy between previous experiments and theory, we demonstrate results from a 1 dimensional simulation of the diffusion equation with a drift current. The results of the simulation show that the net diffusion current in the LPD case of  \cite{Klatt2010} is zero and cannot produce THz emission. We show theoretical analysis of the emission of a dipole under a metal sheet that indicates that the dipole radiation is inhibited, with a similar mechanism to the one in \cite{Drexhage1970}. Therefore, the suppression of dipoles under a semi-infinite metallic sheet creates the anisotropy needed for net current and explains why a net dipole radiates in the geometry of the LPD effect. This claim was investigated experimentally using semi-insulating (SI) and low temperature (LT) grown GaAs substrates and the results are found to be consistent with our hypothesis. Therefore, here we report a novel mechanism of generating THz radiation that is based on the diffusion current created by ultrafast radiation but also uses the inhibition of radiation due to a metal surface. Emitters based on this geometry could have similar or even higher bandwidth to PC antennae without the requirement for an electric bias. \\

 \begin{figure}[htb]
\centering
 \includegraphics[scale=1.1]{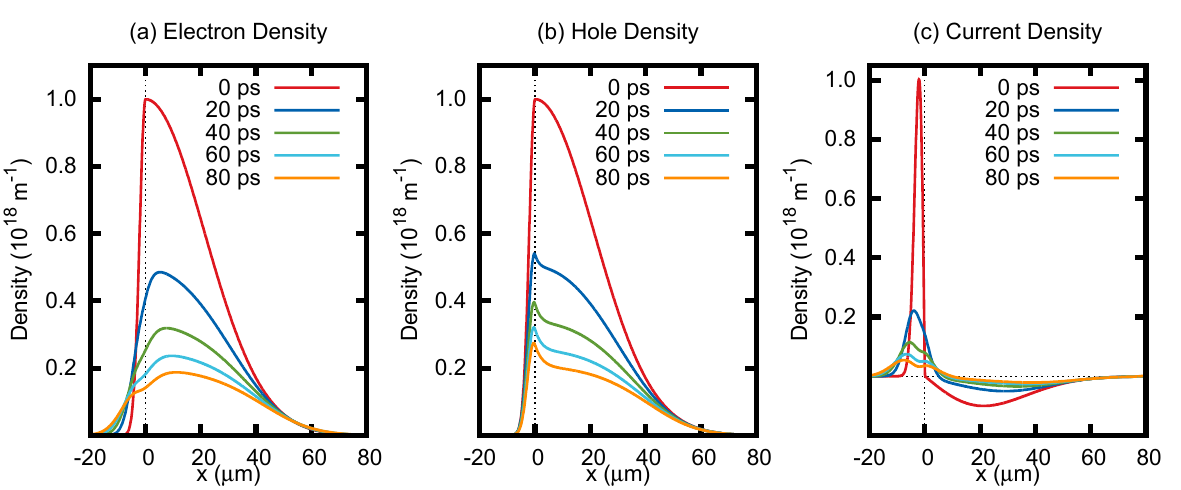}
 \caption{\label{fig:3}(a), (b), (c), representation of the electron density, hole density and current density, respectively. Each curve is at 20 ps intervals within a 80 ps time frame; progression of time corresponds to a fall of the peak concentration or peak current density. }
 \end{figure}

\section{Results of the diffusion equation with drift current}
The geometry of the LPD effect is a semiconductor, such as gallium arsenide, partially masked by a deposited metallic layer, as shown in Fig.\ \ref{fig:1}. The ultrafast laser is focused on the edge of the metal so that the metal obscures half of the beam profile. In order to simulate the carrier diffusion, a 1D model was used similar to the one in \cite{Liu2006,Dekorsy1993}. Assuming carriers in the semiconductor are generated proportional to the light intensity below the metallic mask, the initial concentration of carriers was taken to be a half-Gaussian distribution. As the ultrafast pulse is approximately 100 times shorter than the length of the THz pulse, it is assumed that the carriers are generated instantaneously. The holes are assumed to diffuse with a mobility 20 times lower than that of the electrons \cite{Blakemore1982}. The resulting charge separation produces an electric field, which affects the motion of the carriers. The equation describing the evolution of the electron density, including both electric field and diffusion is,
\begin{equation}
\label{eqn:diffusion}
\frac{\partial n_e}{\partial t}=\mu\frac{\partial}{\partial x}(n_eE)+D\frac{\partial^2n_e}{\partial x^2}-\frac{n_en_h}{\tau_1}-\frac{n_e}{\tau_2}
\end{equation}
where the electron density is $n_e$, the hole densiy is $n_h$ and the electric field is $E$. The mobility, $\mu$ , and the diffusion coefficient, $D$, follow the Einstein relation and the temperature is assumed to be 3000 K for electrons and 300 K for holes \cite{Liu2006,Dekorsy1993}. We assume that the temperature is constant during the simulation. The electron-hole recombination time constant, $\tau_1$, is set at 20 ps for LT and SI-GaAs and the electron-defect recombination time constant, $\tau_2$, is set at 200 fs and used only in the case of LT-GaAs \cite{Gregory2005,Gregory2004,Baker2004}. Equation \ref{eqn:diffusion} is solved numerically for a half-Gaussian with waist of 100 $\mu \textrm{m}$. A mobility of 8500 $\textrm{cm}^2V^{-1}\textrm{s}^{-1}$ was used both for SI and LT-GaAs. Here the results for SI-GaAs are shown, in Fig.\ \ref{fig:3}(a) the temporal evolution of the electron concentration over 80 ps is shown in steps of 20 ps. In this graph the metal sheet shadows the left region from 0 to -20 $\mu\textrm{m}$. The time evolution shows that electrons are annihilated at every step due to recombination and a large flux of electrons towards the left due to diffusion is caused by the initial large gradient. A kink in the electron distribution forms at the metal edge due to the fact that electrons that diffuse to the left recombine at a much lower rate due to the decreased hole density. The hole concentration profiles are shown in Fig.\ \ref{fig:3}(b), diffusion of the holes in this timescale is negligible on account of the lower mobility and lower temperature. A peak forms in the hole concentration at the metal edge because the maximum concentration of electrons moves to the right as a result of diffusion. The current density that is created by this time evolution can be seen in Fig.\ \ref{fig:3}(c), which shows a large positive current spike at the edge of the metal. However, Fig.\ \ref{fig:3}(c) also reveals a negative current in the area of the unmasked semiconductor which shows that there is also diffusion towards the right. This is not obvious in Fig.\ \ref{fig:3}(a) and (b) due to the smooth gradient of the concentration curves. The total current is the integration of these curves of  Fig.\ \ref{fig:3}(c) in space and the result is very small, in comparison to the result of the simulation for a classical PD geometry, but not zero. In order to test if this current is due to diffusion or electric field we ran the simulation only with the effects of diffusion and this resulted to a total current which at any time is exactly zero. Therefore, this small current is due to the effect of the electric field, because the electrons travelling towards the left are attracted by the entire hole population.  The current density due to the electric field is approximately 4 orders of magnitude smaller in comparison to what our simulation predicts for the diffusion current density of the classical PD case \cite{Johnston2002,Liu2006,Gu2002}. In general, the ratio between $E$-field and diffusion generated currents can be estimated to be $\sim r^2/\lambda^2$, where $r$ is a typical length scale for the diffusion and $\lambda$ is the Debye length. The importance of the $E$-field current is dependent on how the Debye length compares with the diffusion distance, in the LPD case $\lambda\gg r$, and diffusion predominates. By comparing the simulation results of the lateral and classical PD emitters we conclude that the current generated from the E-field is negligible and would not be enough to create measurable THz radiation, this explains why the structures depicted in  Fig.\ \ref{fig:2} do not produce measurable THz radiation. \\
\section{Discussion and simulations of dipole radiation}
\begin{figure}[htb]
\centering
 \includegraphics{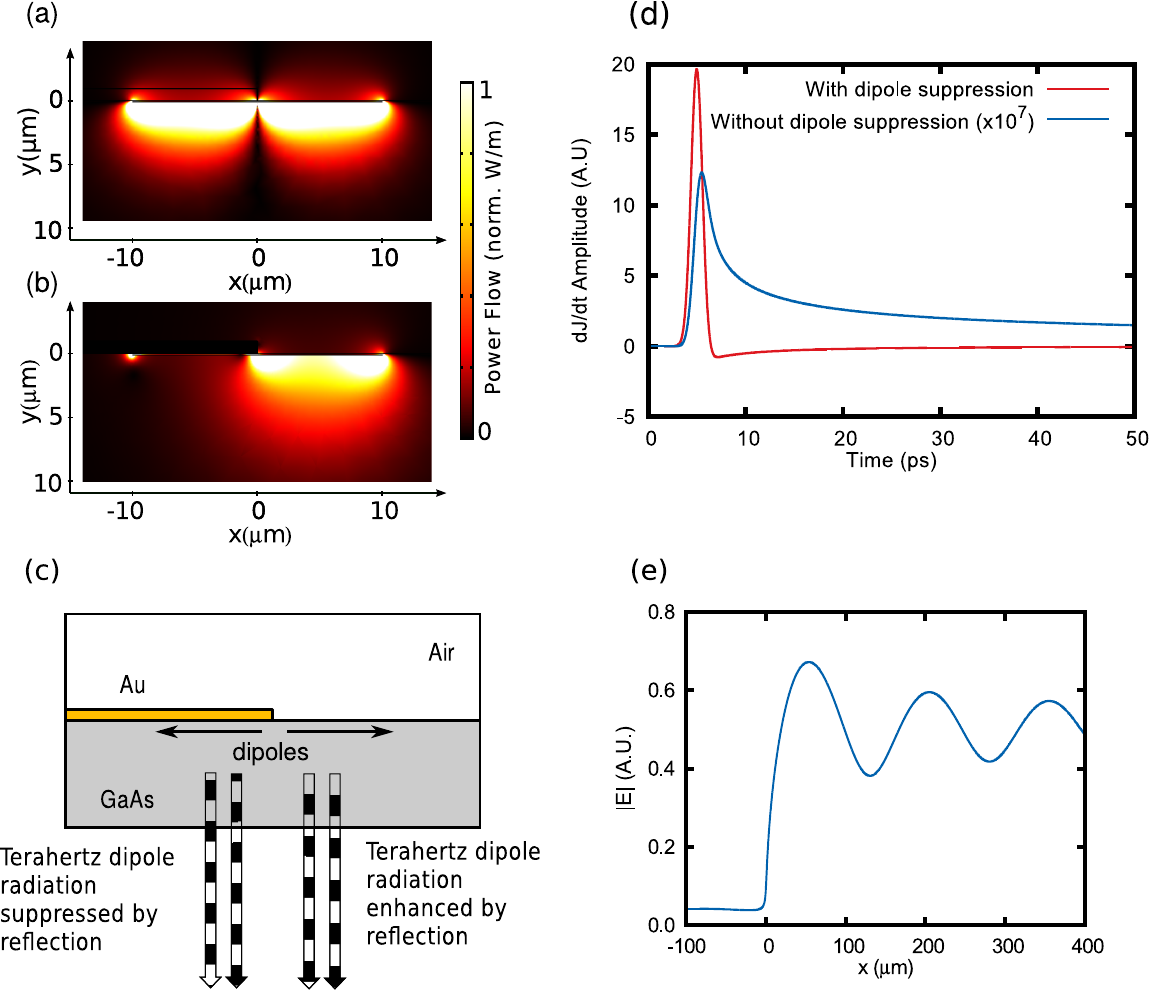}
 \caption{\label{fig:4} (a) and (b) show finite element modelling of a simplified LPD emitter represented by two oscillating currents in anti-phase, 
(a) is modelled without gold masking and (b) with gold masking.(c) Shows the mechanism of radiation in (b) where the dipoles under the metal are suppressed by the reflection of the metal surface. (d) From the numerical solution of Eq. \ref{eqn:diffusion} with a photoexcitation term describing an 1 ps Gaussian optical pulse centered at 5 ps the graph shows the dJ/dt for the total carrier population without taking into account the dipole suppression and the dJ/dt calculated only for the carrier population which is not under metal and therefore free to radiate. (e) Shows a plot of Eq.\ref{eqn:sup} corresponding to the electric field of a dipole radiating
at 2 THz at a distance of 1 $\mu \textrm{m}$ below the metal.}
 \end{figure}

The conclusion from the simulation is that in the case of the LPD geometry diffusion cannot create a net total current even when starting with an asymmetric carrier concentration. Although the simulation is quite simple in assuming instantaneous generation of carriers and stable carrier temperatures, these simplifications do not alter the result that diffusion cannot create a net current. This statement can also be supported with a theoretical argument that the microscopic diffusion current is proportional to $\partial n_e/\partial x$, therefore the total current due to diffusion will be the definite integral $\int\left( \partial n_e/\partial x\right)dx$ from a large negative to positive value. As long as the initial concentration starts from zero and ends to zero this integral will be equal to zero. Furthermore, in diffusion an electron has an equal probability of diffusing in any direction, so the net current due to diffusion must be equal to zero for a large number of electrons.\\
Therefore diffusion cannot be solely responsible for THz radiation for the LPD geometry; another effect must be the cause of the THz emission. Figure.\ \ref{fig:4}(a) shows the emission of two dipoles, as simulated by finite element modelling from two equal oscillating currents in anti-phase, 100 nm under the surface of the semiconductor. In Fig.\ \ref{fig:4}(a) there is no metal, which results in quadrupole radiation that has no component in the direction of $y$ where the THz emission is measured experimentally. In Fig.\ \ref{fig:4}(b) there is a metallic layer (gold) with a refractive index estimated from \cite{Palik1985}. It can be seen that the radiation from the dipole under the metal is completely suppressed; only the dipole in the free area is radiating and of course this gives THz emission in the $y$-direction of the receiver and the mechanism is depicted in Fig.\ \ref{fig:4}(c). The suppression of radiation happens due to the long wavelength of the THz radiation in relation to the distance between the dipole and metal. The radiation that is reflected by the surface of the metal acquires a $\pi$-phase shift in relation to the non-reflected radiation \cite{Doi1997,Yasuda2008}. Interference between the non-reflected and reflected radiation is destructive and causes the dipole under the metal to be suppressed, and no emission to be generated in the $y$ direction  \cite{Drexhage1970}. As an estimation of the radiated THz field we have calculated the rate of change of total current given by the numerical solution of Eq. \ref{eqn:diffusion} with the addition of a photoexcitation term describing an 1 ps-FWHM-Gaussian optical pulse centered at 5 ps to generate the carriers. The results are shown in Fig.\ \ref{fig:4}(d) that shows the rate of change of current for the whole carrier population (representing the case that the whole carrier population could radiate), and the rate of change of current for only the carrier population which is free to radiate (not under the metal). The results show that the predicted THz radiation taking into account the dipole suppression is the dominant mechanism and is responsible for breaking the symmetry in the LPD geometry. \\

We analytically studied the radiation of a dipole under a semi-infinite metallic surface using \cite{Morse1953}. The problem is considered in 2 dimensions, over a cross-section of the emitter. The surface of the semiconductor and the metal is in the $x$-direction and the THz wave is generated in the $y$-direction, as shown in Fig.\ \ref{fig:1}(a). An expression for the radiated electric field from an oscillating dipole under the surface is:
\begin{eqnarray}
\label{eqn:sup}
E=\frac{1}{2\sqrt{i\pi}}\left[e^{iky_0}\Phi\left(\sqrt{k\left(r_0+y_0\right)}\right)\right.\nonumber\\*
\left.-e^{-iky_0}\Phi\left(\pm\sqrt{k\left(r_0-y_0\right)}\right)\right]
\end{eqnarray}
Where $r_0$ is the distance of the dipole from the edge of the metal sheet and $y_0$ is the distance in the vertical direction and $k$ is the wavenumber. $\Phi$ is the Fresnel integral defined by $\Phi(z)=\int^{z}_{-\infty}e^{it^2}dt$. The minus sign in the radical applies when $x$ is positive and the positive sign when $x$ is negative. The derivation is done in free space which underestimates the amount of suppression and a perfect metal has been assumed.\\
 \begin{figure}[htb]
\centering
 \includegraphics{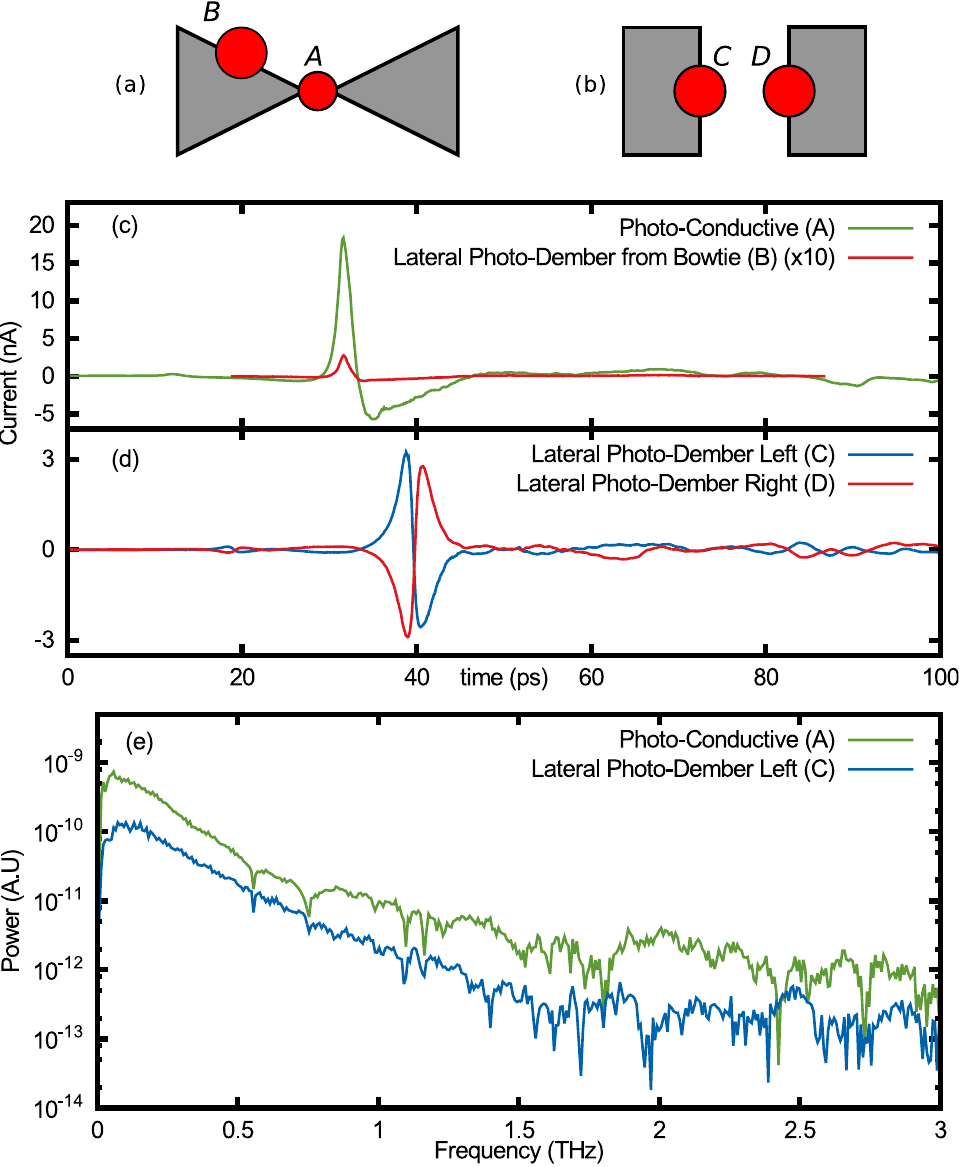}
 \caption{\label{fig:5}Diagram showing the placement of the laser beam on (a) the bowtie PC emitter and (b) the LPD emitter. (c) Shows the current recorded in the time domain for a PC antenna used as a PC and a LPD emitter. (d) shows
THz emission at two opposite boundaries to observe the predicted sign change. (e) shows the Fourier transform of (a) and (b) demonstrating comparable bandwidth between the PC and LPD emitter. The experiment was performed in ambient atmosphere.}
 \end{figure}

Equation \ref{eqn:sup} is plotted in Fig.\ \ref{fig:4}(e) using \cite{Abramowitz1972}, where it shows the dependence of the radiation in the $y$ direction as a function of the position of the dipole. As expected, when under the metal there is suppression, whereas when the dipole is out of the metallic region it emits. The amount of suppression depends on how close the dipole is to the metal and the oscillation frequency. Here the dipole separation from the metal is set at 1 $\mu \textrm{m}$ as an approximation of the absorption length of GaAs at a wavelength of 800 nm \cite{Liu2006} and a frequency of 2 THz has been chosen as a typical emission frequency. The emitted radiation drops close to zero, even when the dipole is just below the edge of the metal. The oscillations in the strength of radiation reveal interference phenomena, however in the case of experiment, where a wide spread of dipoles is present, the oscillations are not expected to be noticeable. \\
Therefore, the surface in the classical photo-Dember effect \cite{Johnston2002,Liu2006,Gu2002} breaks the symmetry of the diffusion process because carriers are not permitted to diffuse out of the surface. Thus a net diffusion current is created that emits a THz wave. In the case of the LPD there is no break of symmetry in the diffusion process, carriers diffuse both ways and thus the net current is zero (or very small if the electric field contribution is accounted for). The break of the symmetry is created because the metal suppresses the dipole radiation of the carriers underneath it and only the carrier population that is not under the metal and mostly diffuses towards the opposite direction is free to radiate. \\
The array of emitters depicted in Fig.\ \ref{fig:2} were designed, following \cite{Klatt2010}, to create asymmetric concentration profiles. However, it has been shown here theoretically that diffusion cannot create a net current in the LPD geometry and electric field contributions are weak, therefore this explains why the array did not work as a THz emitter. Conversely, the multiple emitters presented in \cite{Klatt2010} work as THz emitters. This is possibly due to the use of metallic wedges with thickness comparable to the skin depth of gold in the THz spectrum. The variation of thickness introduces an asymmetry; because part of the dipole population under the metallic wedges is not totally suppressed, emission of THz radiation is possible.  \\
\section{Experimental verification}
A theory of how the LPD emitter works is proposed here and it is straightforward to experimentally validate. The hypothesis in  \cite{Klatt2010} states that the discontinuity in the carrier distribution will create a dipole directed towards the metallic region while our theoretical model predicts the formation of a net dipole in the opposite direction due to the suppression of dipoles beneath the metal. In THz-time domain spectroscopy the direction (sign) of the E-field can be directly measured and compared between experiments. Experimentally the THz radiation generated is collimated and focused with two parabolic mirrors onto an LT-GaAs receiver. A 5 $\mu$m-gap bowtie PC antenna was used, as shown in Fig. \ref{fig:5}(a) with the laser focused on top of position A. The polarity of the THz emission was mapped with the direction of current by biasing the PC antenna in opposite polarities and one of the measurements is shown in Fig.\ \ref{fig:5}(c). The same PC antenna (LT-GaAs) was then disconnected from the bias, and translated across to the area B, shown in Fig. \ref{fig:5}(a), where a metallic edge was used as a LPD emitter. The polarity of the THz waveform indicated that the current was flowing as expected in our argument where the radiating dipole is in the non-masked region of the antenna and the THz waveform is also shown in Fig. \ref{fig:5}(c). The LT-GaAs PC emitter was then replaced with a dedicated LT-GaAs-LPD emitter and we measured the LPD effect on two opposite edges of metal strips, areas C and D in Fig. \ref{fig:5}(b); the beam waist at the focus was approximately 60 $\mu$m. The results are shown in Fig.\ \ref{fig:5}(d), where we note the expected polarity change between opposite edges. In Fig.\ \ref{fig:5}(e) the spectra of PC and LPD antennae are illustrated, showing similar bandwidths produced from the LPD and PC emitter.  The spectra were obtained in ambient atmosphere and show water absorption features.\\
SI-GaAs samples have also been used with similar results, however, the LT-GaAs samples had much larger bandwidth and signal in comparison to the SI-GaAs. The LT-GaAs emitter was easier to saturate and thus a larger area was illuminated, in relation to SI-GaAs but further theoretical and experimental investigations are needed to investigate these differences in performance between LT and SI-GaAs. \\
\section{Conclusion}
In conclusion, in a partially metal-masked semiconductor, illuminated with ultrafast NIR radiation, carrier diffusion and carrier recombination alone cannot account for the observed THz radiation. We propose a new theory for the THz emission due to dipole radiation suppression from the metal mask and experimental evidence that supports our model for THz emission. This type of emitter does not suffer from the lifetime issues of biased PC antennae as there is no electrical bias requirement. Furthermore, emitters based on lateral diffusion currents and dipole radiation suppression are simple to fabricate, opening up possibilities for easier THz integration and interfacing with other elements. This concept gives rise to design proposals for a series of emitters which would give similar performance to a PC antena, which is currently the standard THz emitter. 
\end{document}